\documentclass[9pt,twocolumn,twoside]{pnas-new}

\templatetype{pnasresearcharticle}
\setboolean{displaywatermark}{false}
\usepackage{expl3}
\usepackage[version=4]{mhchem}

\usepackage[detect-all]{siunitx}			 
\DeclareSIUnit\kbt{k_{B}T}
\definecolor{myblue}{rgb}{0.0,0.57,0.81}

\def\y{y}

\newcommand{\kn}{\kappa_{n}}
\newcommand{\vn}{e_n}

\newcommand{\kc}{\kappa_{c}}
\newcommand{\vc}{e_c}
\newcommand{\deltav}{\Delta e}
\newcommand{\vzero}{e_0}

\newcommand{\kb}{k_{\mbox{\tiny B}}}
\newcommand{\vsnare}{E_{\mbox{\tiny snare}}}

\newcommand{\fsnare}{f_{\mbox{\tiny snare}}}

\newcommand{\vfusion}{E_{\mbox{\tiny fusion}}}

\newcommand{\vf}{e_f}
\newcommand{\yf}{y_f}
\newcommand{\sigmaf}{\sigma_f}
\newcommand{\taufusionbar}{\bar \tau_{\mbox{\tiny fusion}}}
\newcommand{\taufusion}{\tau_{\mbox{\tiny fusion}}}

\newcommand{\rmd}{{\mbox{d}}}

\title{SNARE~machinery~is~optimized~for~ultra-fast~fusion}

\author[a]{Fabio Manca}
\author[a]{Frederic Pincet}
\author[b]{Lev Truskinovsky}
\author[c,d]{James E. Rothman}
\author[a]{Lionel Foret} 
\author[e,1]{Matthieu Caruel}

\affil[a]{ Laboratoire de Physique de l’Ecole Normale Supérieure,  Ecole Normale Supérieure (ENS) - CNRS - Université PSL - Sorbonne Université - Université Paris-Diderot - Sorbonne Paris Cité, 75005 Paris, France}
\affil[b]{Physique et Mécanique des Milieux Hétérogènes (PMMH), Ecole Supérieure de Physique et de Chimie Industrielles (ESPCI) -  CNRS - Université PSL, 75231 Paris cedex 05, France.}
\affil[c]{Department of Cell Biology, Yale University, New Haven, CT 06520, USA; }
\affil[d]{Department of Experimental Epilepsy, Institute of Neurology, University College London, London, United Kingdom}
\affil[e]{Laboratoire Modélisation et Simulation Multi-Echelle (MSME), Université Paris-Est, CNRS, 94010 Créteil Cedex, France}

\leadauthor{Manca} 

\significancestatement{
We propose a new mechanistic description for the fusion of a synaptic vesicle with a target membrane executed by a team of  zippering SNAREpins.
In the context of neurotransmitters release, this process naturally decomposes in two steps whose rates depend on the number of SNAREpins \( N \).
The first step is the synchronized escape from the metastable half-zippered state, which get exponentially more sluggish as \( N \) increases.
The second step is the fusion of the two closely tethered membranes, which is accelerated exponentially by an increase of \( N \).
The tradeoff between these two antagonist trends results in a sharply  optimal number of SNAREpins \( N=\numrange{3}{6}\) which ensure fusion at the physiological sub-\si{\milli\second} timescale.
}

\authorcontributions{M.C. L.F.  and L.T. designed research; F.M. performed research; F.M. analyzed data; F.M., F.P., L.T., J.E.R., L.F., and M.C. interpreted the results and wrote the paper}
\authordeclaration{The authors declare no conflict of interest.}
\correspondingauthor{\textsuperscript{1}To whom correspondence should be addressed. E-mail: matthieu.caruel@u-pec.fr}


\begin{abstract}
		SNARE proteins zipper to form SNAREpins that power vesicle fusion with target membranes in a variety of biological processes.
	A single SNAREpin takes about 1 second to fuse two bilayers, yet a handful can ensure release of neurotransmitters from synaptic vesicles much faster, in a 10th of a millisecond.  
	We propose that, similar to the case of muscle myosins,  the ultrafast fusion results from cooperative action of many SNAREpins. 
	The coupling originates from mechanical interactions induced by confining scaffolds. Each  SNAREpin is known to have enough energy to overcome the  fusion barrier of \SIrange{25}{35}{\kbt},  however, the fusion barrier only becomes relevant when the SNAREpins are nearly completely zippered and from this state each SNAREpin can deliver only a small fraction  of this energy as mechanical work.
Therefore they have to act cooperatively and   we show that at least 3 of them are needed to ensure fusion in less than a millisecond.
However, to reach the pre-fusion state collectively,
starting from the experimentally observed  half-zippered metastable state,  the SNAREpins have to mechanically synchronize which takes exponentially longer time as the number of SNAREpins increases.
Incorporating this somewhat counter-intuitive idea in a simple coarse grained model results in the novel prediction that there should be an optimum number of SNAREpins for sub-\si{\milli\second} fusion: \numrange{3}{6} over a wide range of parameters. Interestingly, \emph{in situ} cryo-electron microscope tomography has very recently shown that exactly six SNAREpins participate in the fusion of each synaptic vesicle. This number is in the range  predicted by our theory.

\end{abstract}

\begin{document}

\verticaladjustment{-2pt}

\maketitle
\thispagestyle{firststyle}
\ifthenelse{\boolean{shortarticle}}{\ifthenelse{\boolean{singlecolumn}}{\abscontentformatted}{\abscontent}}{}

\section{Introduction} 
\label{sec:introduction}

\dropcap{P}rotein transport within cells relies heavily on membrane-enveloped vesicles that ferry packets of enclosed cargo  \cite{Ivanov:2008gw,Vassilieva:2008hk,Jahn:2012kq,Sudhof:2009cw}.
The content of the  vesicles is released \emph{via} their fusion with target membranes.
This transition is impeded by repulsive forces acting when the distance between the membranes is in the range of \SI{\sim 1}{\nano\meter}.
The encountered energy barrier is of the order of \SI{30}{\kbt}, implying that spontaneous fusion would take minutes, which is not fast enough in most biological situations \cite{Rand:1989iy,Leckband:2001bq,Ryham:2016gu,FrancoisMartin:2017db}.
For this reason, the process is assisted by SNARE proteins (soluble N-ethylmaleimide-sensitive factor attachment protein receptors, SNAREpins) whose conformational change (zippering) exerts forces pulling the vesicle membrane towards the target membranes. 

While the total free energy change associated with the zippering process is of the order of \(  \SI{\sim70}{\kbt} \) \cite{Zhang:2017fm}, most of this energy is consumed as the SNAREpins bring the membranes into close apposition.
Biologically, the initial assembly prior to fusion provides compartmental specificity (pairing the correct SNAREs together) and
allows for temporal regulation (clamping).
Terminal zippering is then the  process which utilizes the remaining  energy  for bilayer fusion at the small (\(\SIrange{\sim1}{2}{\nano\meter} \)) separations where the repulsive forces become relevant.
Recent studies suggest that each SNAREpin can deliver only about \SI{5}{\kbt} of mechanical work at this stage \cite{Gao:2012gu,Zhang:2008bp}, which  explains why it takes about 1 second for a single SNAREpin to fuse two bilayers \cite{Xu:2016jda,Domanska:2009cw}.

It is known however that the release of neurotransmitters from synaptic vesicle occurring at nerve endings happens considerably faster, in a 10th of a \si{\milli\second} as is necessary to keep pace with action potentials, and ensure synchronous release \cite{Sudhof:2004gh,Sudhof:2009cw,Camacho:2017gd,Liu:2016cf,Li:2016isa,Krishnakumar:2015de}.
A widely accepted explanation for this remarkable difference in time scales is  that multiple SNAREpins would need to cooperate to accelerate fusion after being   synchronously released from a clamped state.
There have been indirect indications that the number of SNAREpins necessary to achieve a sub-millisecond fusion may be relatively small, ranging from 2 to 6 \cite{Hua:2001bb,Sinha:2011gxa,Mohrmann:2010eg}.
Very recently, cryo-electron microscope tomography of synaptic vesicles \emph{in situ} revealed an underlying 6-fold symmetry suggesting that exactly 6 SNAREpins are involved in such processes  \cite{Li:2019}.  

How so few co-operating SNAREpins manage to accelerate fusion ten thousand times  (from \( \SI{\sim1}{\second} \) to \(\SI{\sim0.1}{\milli\second} \))   has been a mystery.
Previous modeling attempts   have suggested that more than 16  SNAREpins would be required  \cite{Mostafavi:2017gd,McDargh:2018jd}.
Here we show  that the key to understand how  only a  few  SNAREpins can achieve such rapid fusion  is the simple fact that they are mechanically coupled through effectively rigid common membranes.
The account of such mechanical coupling leads to a striking prediction  that the  number of SNAREpins must be highly constrained to ensure  sub-millisecond release  of neurotransmitters. Quite remarkably, the predicted  optimal range,     \numrange{3}{6}, is in excellent agreement with most recent experimental results \cite{Li:2019}.  

We draw a fundamental analogy between the collective zippering of the SNAREpins and the  \emph{power-stroke} in a bundle of elastically coupled muscle myosin II proteins which is known to also take place at \SI{1}{\milli\second} timescale.
Building upon the seminal theory of the myosin power-stroke proposed by Huxley and Simmons \cite{Huxley_1971,Caruel:2013jw,Caruel:2016kw,Caruel:2017eba}, we model the fusion machinery as a mechanical system where the SNAREpins are represented as snap-springs interacting  through supporting membranes.
The implied  bi-stability is supported by recent experiments showing the presence of a metastable half-zipped state \cite{Gao:2012gu,Zhang:2008bp}.

The theoretical approach developed in this paper highlights  the essential role of mechanical coupling among proteins undergoing conformational changes in ensuring swift, highly  synchronized mechanical response.  This is likely a general biological principle  \cite{Caruel:2016kw,Caruel:2018jg}.

\begin{figure}[tbp]
	\centering
	\includegraphics[scale=0.22]{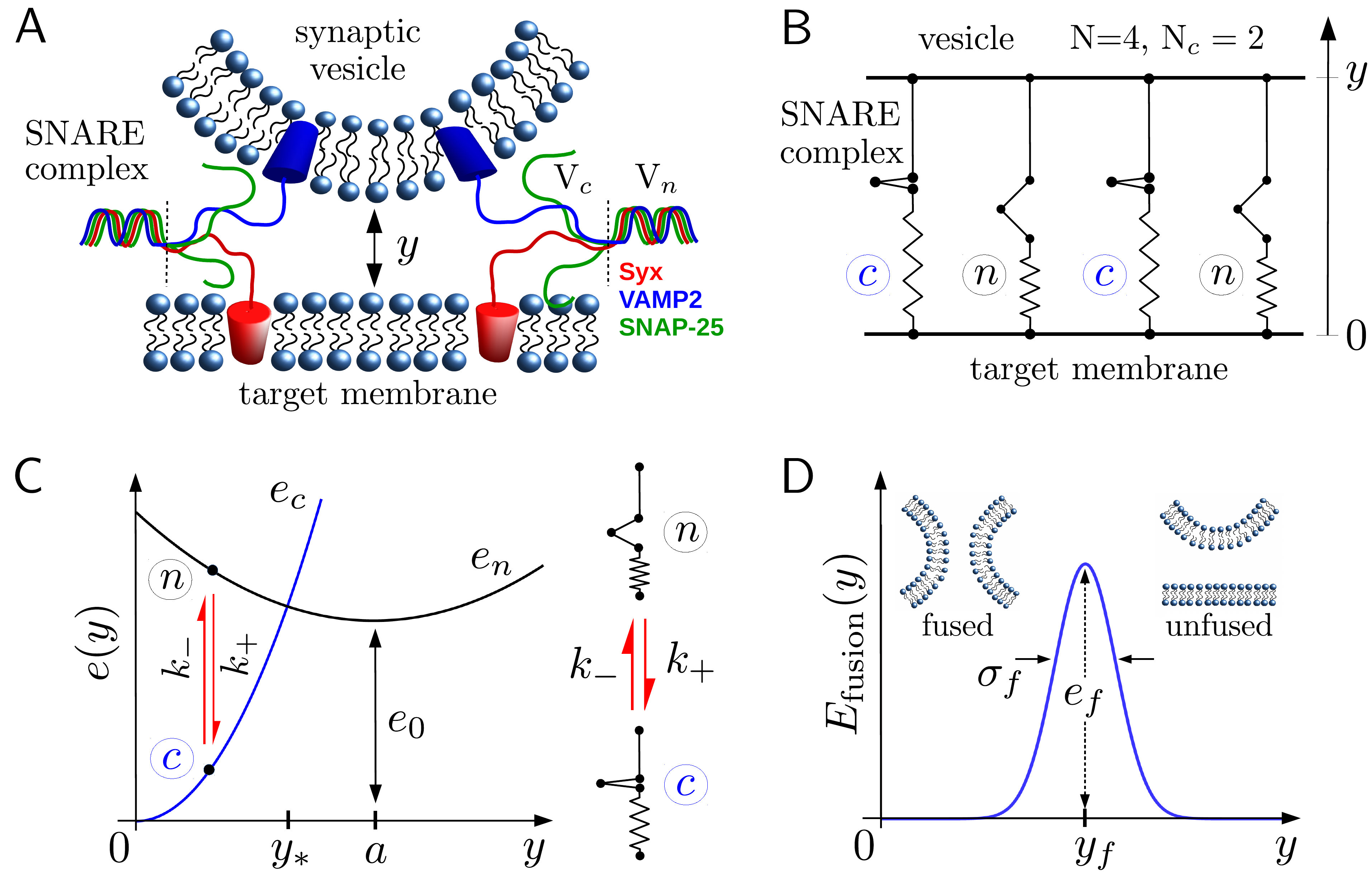}
	\caption{
	The fusion machinery. (A) schematic of the two membranes with two attached SNAREpins. (B) Mechanical model with \( N=4 \) SNAREpins in parallel bridging the two membranes separated by the distance \( y \); two SNAREpins are in state \( c \) and two are in state \( n \) so \( N_{c}=2 \). (C) Model of a single SNAREpin. (D) Fusion energy landscape. See Table~\ref{tab:parameters} for the complete list of parameter values.
	}
	\label{fig:model}
\end{figure}

 \begin{figure*}[!htbp]
 	\centering
 	\includegraphics[scale=0.34]{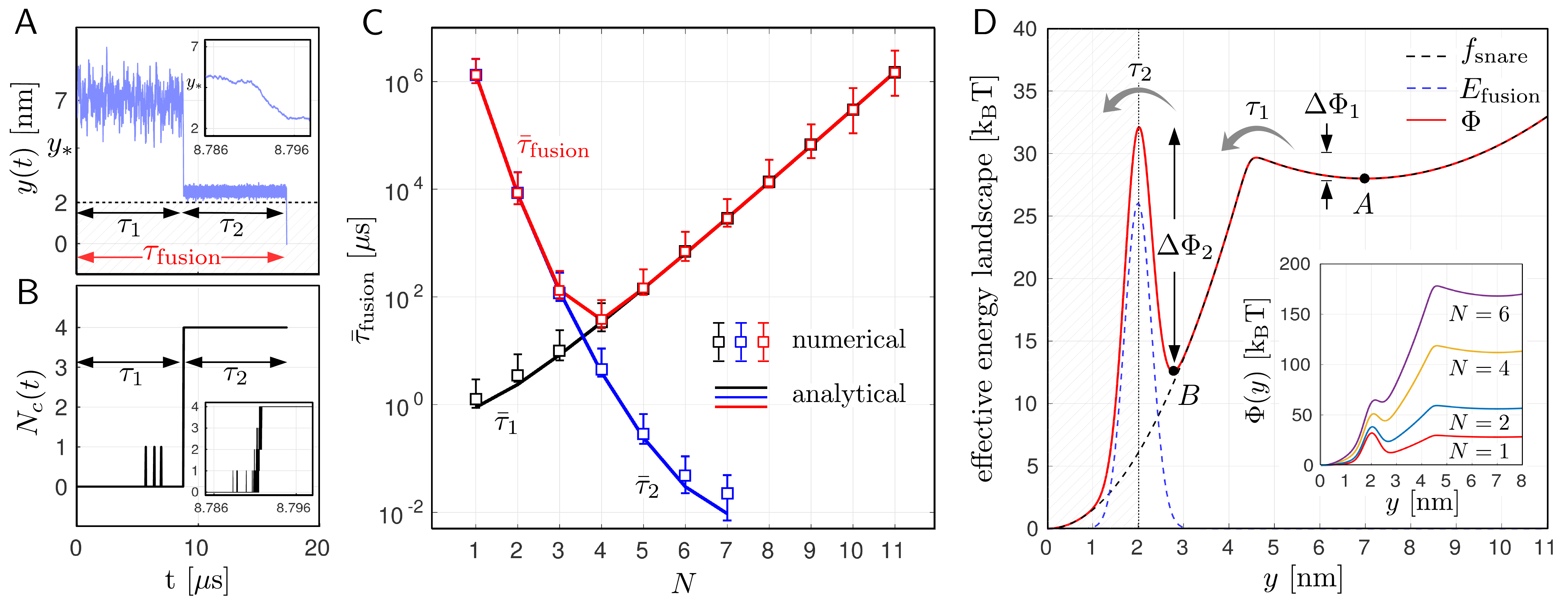}
 	\caption{
 	Main results. [(A) and (B)] typical stochastic trajectories of the inter-membranes distance \( y \) (A) and the number \( N_{c}(t) \) of SNAREpins in state \( c \) (B) obtained from the numerical simulation. (C) Average of the waiting times \( \tau_{1} \) (black), \( \tau_{2} \) (blue) and \( \taufusion=\tau_{1}+\tau_{2} \) (red) obtained from the numerical simulations (symbols) and from our effective chemical model (lines). (D) Effective free energy landscape \( \Phi \) showing the three stages of fusion and the associated transition rates. Parameters are listed in Table~\ref{tab:parameters}.
 	}
 	\label{fig:results}
 \end{figure*}
\section{Fusion Machinery} 
\label{sec:fusion_machinerey}

The goal of the model is to describe the dynamic coupling between the individual SNAREpins zippering  and to study the associated evolution of the distance between the vesicle and a the target membrane.
The assembled SNARE machinery  is represented as a bundle of $N$  parallel SNAREpins bridging the two membranes, separated by the distance \( y \),  see Fig.~\ref{fig:model}[(A) and (B)].
We assume that irreversible fusion occurs when this distance reaches a critical value \( y_{f} \).
  The characteristic length associated with the deformation of the membranes generated by a zippering SNAREpin is large compared to the typical size of the SNARE bundle, see Materials and Methods \ref{sub:rigid_membrane_assumption}. Hence, the membranes can be viewed as two rigid backbones cross-linked by \( N \) identically stretched SNAREpins.

\paragraph{Single SNAREpin as a bi-stable snap-spring}The experimental work conducted in Refs.~\cite{Sutton:1998cw,Gao:2012gu,Zhang:2008bp}, suggests that a single SNAREpin can switch randomly between two metastable conformations: $n$ (half-zippered) when only the N-terminal domain of the SNAREs is zippered and $c$ (fully zippered),  when both the C-terminal and the linker domain are zippered.
To describe this process, we assume that the half-to fully zippered transition in a SNARE complex  is similar to the pre- to post-power-stroke conformational change in a myosin motor, \cite{Huxley_1971,Caruel:2016kw}.

Suppose that each SNAREpin is equipped with an internal spin type degree of freedom characterizing  the state of the protein, \( n \) or \( c \).
We denote by \( a \) the amount of shortening resulting from the \( n\to c \) transition in the absence of external load,
and by \( \vzero \) the energy difference between the two states.
This parameter can be interpreted as the typical amount of mechanical work necessary to force the \( c\to n \) transition (partial unzipping), see Fig.~\ref{fig:model}(C).

When the SNAREs are bound to the membranes, we assume that the rates $k_{+}(y)$---associated with the \( n\to c \) transition---and \( k_{-} \)---associated with the \( c\to  n\) transition---depend on the mechanical load induced by the variations of the inter-membrane distance.
To specify this dependence, both states are assumed to be  ``elastic'' in the sense  that they  exist, as phases, over an extended range of separations $y$, due to elongations of the zippered and unzippered SNARE residues, internal bonds rearrangement etc..., see Fig.~\ref{fig:model}(C). 
For simplicity we assume that the deformations remains in the elastic regime so that states \( n,c \) can be associated with quadratic energies \( e_{n,c}(y) \), with minima located \( y=\{0,a\} \), and with the lumped stiffnesses \( \kappa_{n,c} \). 
The transitions rates are defined so that, for a given separation, they favor the state with the lowest energy and verify detailed balance. For the detailed expressions of \( e_{n,c} \) and \( k_{\pm} \), see Materials and Methods \ref{sub:model_of_a_single_snarepin}.

\paragraph{Dynamics of the fusion machinery}
The parallel arrangement of the SNAREs implies that the conformational state of the bundle is fully characterized by $N_c$, the number of SNAREpins in state $c$. 
This variable evolves according to the stochastic equation $N_c(t+dt) =N_c(t)+\{1,-1,0\}$,   with the outcomes \( \{1,-1,0\} \), characterized  by the probabilities $W_{+1}(y,N_c)=(N - N_c)\,k_+(y)dt $,  $W_{-1}(y,N_c)=N_c\,k_-(y)dt $ and $W_{0}=1-W_{+1}-W_{-1}$.
While  the SNAREpins can switch independently, the transition rates \( k_\pm \) are functions of  the  collective variable  $y$, whose dynamics in turn depends on \( N_{c} \).
 \begin{table}[tbp]
 \centering
 \caption{Physical parameters adopted in the model and references}
 \begin{tabular}{lp{0.5cm}ccp{1cm}}
 \toprule
 Parameter & Symbol & Value & Units & Reference \\
 \midrule
 Zipping distance & $a$ & \num{7} & \si{\nano\metre} & \cite{Gao:2012gu}\\
 Energy bias & $\vzero$ & 28 & \si{\kbt} & \cite{Gao:2012gu}\\
 Fully zipped stiffness & $\kc$ & \num{12} & \si{\pico\newton\per\nano\metre} & SI** \\
 Half-zipped stiffness & $\kn$ & \num{2.5} & \si{\pico\newton\per\nano\metre} & SI  \\ 
 Maximum zippering rate & $k$ & \num{1} & \si{\mega\hertz} & \cite{Gao:2012gu}\\
 \midrule
 Drag coefficient & $\eta$ & \num{3.8e-07} & \si{\newton\second\per\meter} &  \\
 \midrule
 FB* position & $\yf$ & \num{2} & \si{\nano\metre} & \cite{Evans:1991fz, Rand:1989iy}\\
 FB width & $\sigmaf$ & \num{0.3} & \si{\nano\metre} & \cite{Leckband:2001bq,Donaldson:2011cn} \\
 FB height & $\vf$ & \num{26} & \si{\kbt} & \cite{Ryham:2016gu,FrancoisMartin:2017db}\\
 \bottomrule
 \label{tab:parameters}
 \vspace{-0.5cm}
 \end{tabular}
 \flushleft \vspace{-0.3cm}
 \addtabletext{*FB: Fusion Barrier; **SI: SI Appendix; \( \SI{1}{\kbt}\approx\SI{4}{\zepto\joule} \)
 }
 \end{table}  
 
To specify the coupling between the two degrees of freedom \( N_{c} \) and \( y \), and thereby formulate the complete model of the fusion process, we first recall that the  motion of the vesicle in the overdamped regime results from the balance between the force applied by the \( N \) SNAREpins, the membrane repulsion and the viscous drag.
Taking into account the thermal fluctuations, this force balance translates into the stochastic equation, 
\begin{equation}
	\label{eq:dynamics_y}
		\eta \dot \y  =    -\frac{\partial}{\partial y} (\vsnare+\vfusion)  + \sqrt{2\eta\,\kb T}~\xi(t),
\end{equation}
where $\xi(t)$ is a standard  white noise, and  $\eta$ is a drag  coefficient  representing the friction opposing the motion of the vesicle. 
At a given \( y \), the force applied by the bundle derives from the sum of individual SNAREpin energies $\vsnare(y,N_c) = N_c\, e_c(y) + (N-N_c)\, e_n(y)$.
Finally, the inter-membrane repulsion, due to short-range   forces  between the two membranes, is schematically modeled by a Gaussian energy barrier \cite{Rand:1989iy} $\vfusion(y)= \vf \exp[-(y-\yf)^2/(2\sigmaf^2)]$, where $\yf$ is the critical separation,  while $\vf$  and $\sigmaf$ are the height and the width of the barrier, respectively, see Fig.~\ref{fig:model}(D).
  The ensuing dynamics of the  system unfolds in the space of two variables:  the continuous  one, $y(t)$, and the integer-valued one, $N_c(t)$.
  The associated energy landscape has a multiwell structure that accounts for the configurational states of \( N \) individuals.
  The response is governed by the two stochastic equations, for which initials conditions still need to be specified.
 We consider the initial state \( N_{c}(t=0)=0 \), and  \( y(t=0) = a \), which corresponds to the configuration where the SNAREpins are at the bottom of the energy well describing  state $n$. 
 This configuration characterizes the system immediately after the calcium induced collapse of Synaptotagmin which triggers the full zippering of the SNAREs \cite{Sudhof:2013fw}. This point is discussed in more details in Section~\ref{sec:discussion}.

 
\paragraph{Model parameters}
The model is calibrated as follows, see Table~\ref{tab:parameters} and SI Appendix~(A) for further details. The mechanical parameters characterizing a single SNAREpin, \( a \), \( \vzero \), \( \kappa_{n} \) and \( \kappa_{c} \), are determined by using our model to reproduce the experimental results obtained from stretching tests with optical tweezers, see Refs.~\cite{Gao:2012gu,Zhang:2017fm}.
The energy bias \( a \) and \( \vzero \) are chosen to be compatible with the results obtained from these studies.
The procedure used to estimate the stiffnesses \( \kappa_{n,c} \) is more complex and explained in details in SI Appendix.
The value of the rate \( k \), is fixed in accordance with estimates from Refs.~\cite{Gao:2012gu,Yang:2003ib}.
The drag coefficient is computed using the Stokes formula $\eta = 6\pi \mu R$, where $R=\SI{20}{\nano\meter}$ is the vesicle radius, and $\mu=\SI{e-03}{\pascal\second}$ the fluid viscosity.
 The corresponding characteristic timescale is $\tau_{\eta} = \eta a^2 / (\kb T) \approx \SI{4.5}{\micro\second}$.

The values of the parameters \( y_{f} \), \( \sigma_{f} \) and \( \vf \) are chosen to be compatible with the current literature \cite{Ryham:2016gu,Lentz:2009jv, Zhang:2008bp,Cohen:2004ds,Mostafavi:2017gd,McDargh:2018jd,FrancoisMartin:2017db,markvoort2011,finkelstein1986,grafmuller2007,grafmuller2009,lee2008,Xu:2016jda}.
In particular, values of \( \vf \) between 26 and \SI{34}{\kbt} have been reported for various types of lipids. We chose \SI{26}{\kbt} [POPC\footnote{1-palmitoyl-2-oleoyl-sn-glycero-3-phosphocholine lipid}, see Ref.~\cite{FrancoisMartin:2017db}] which leads to a single SNAREpin average fusion time of \SI{1}{\second}.


\section{Results} 
\label{sec:results}

 \paragraph{Numerical simulations} Typical stochastic trajectories, \( y(t) \) and  \( N_{c}(t) \) obtained from numerical simulations, are shown in  Fig.~\ref{fig:results}[(A) and (B)]. They indicate that the fusion process can be decomposed into two stages, characterized by the times \( \tau_{1} \) and \( \tau_{2} \). 
 During the first stage, the system remains in its initial configuration \( (y\simeq a, N_{c}=0) \) with only isolated  \( n\to c\to n \) transitions. After a   time \(  \tau_{1} \), the   inter-membrane distance drops abruptly to \( y\simeq \SI{2.5}{\nano\metre}\), while all the SNAREpins collectively switch from state \( n \) to state \( c \).  
 The inserts in Fig.~\ref{fig:results}[(A) and (B)] show that this transition occurs within \SI{10}{\nano\second} after the inter-membrane distance has reached the value \( y=\y_* \); the irreversible collective zippering  itself (\( N_{c}=0\to N_{c}=4 \)) lasting about \SI{1}{\nano\second}.
After the synchronized \( n\to c \) transition the inter-membrane distance remains above the threshold \(y= \yf \) for a time \( \tau_2 \) before fusion. The duration of the whole process is therefore \( \taufusion=\tau_{1}+\tau_{2} \).

The mean timescales \( \bar{\tau}_{1} \) and \( \bar{\tau}_{2} \) (obtained by averaging \num{e03} stochastic trajectories) are represented as functions of the number of SNAREpins in Fig.~\ref{fig:results}(C) on a semi-logarithmic scale. Observe  that  \( \bar{\tau}_{1} \) increases exponentially with \( N \) and   \( \bar{\tau}_{2} \) decreases exponentially with \( N \).  
These antagonistic  \( N \)-dependencies result in  the average fusion time \( \taufusionbar=\bar\tau_{1}+\bar\tau_{2} \) exhibiting a remarkably sharp minimum, see Fig.~\ref{fig:results}[(C), red].
With the set of parameters values reported in Table~\ref{tab:parameters}, this minimum is attained at \( N_{*}=4 \) and is associated with a fusion timescale of \SI{\sim 100}{\micro\second}. In addition we obtain a fusion time of the order of \SI{1}{\second} for a single SNAREpin.
Both values are consistent with \emph{in vitro}  \cite{Xu:2016jda} and \emph{in vivo} \cite{Meinrenken:2002dc,Sudhof:2004gh} experimental measurements.

\paragraph{Fusion as a two-stage  reaction} To elucidate the mechanism of fusion in two stages, we present here a `toy' model where the whole process is recast as two successive reactions 
\begin{equation}
\label{eq:chemical_reaction}
\ce{unfused <=>>[$k_{1}$][$k_{-1}$] IS ->[$k_{2}$] fused},
\end{equation}
where IS stands for an intermediate state whose characteristics depend on the mechanical properties of the zippered SNAREpins. In this representation,  the fusion is viewed as the outcome  of two distinct sub-steps: the collective zippering and the topological membrane merger. 

To justify such model reduction, we assume that the timescale of the  \( n\rightleftharpoons c  \) transition is negligible compared to the timescale describing the relaxation of the vesicle position.
In the corresponding limit ($kt_{\eta}\gg 1 $) \eqref{eq:dynamics_y} can be averaged with respect to the equilibrium distribution of the variable \( N_{c}(t) \) [see Materials and Methods~\ref{sub:adiabatic_elimination_of_the_variable_boldsymbol_n__c}], so the original system reduces to the one-dimensional stochastic equation
\begin{equation}
	\label{eq:effective_langevin}
	\eta\dot{y} = -\,\frac{\rm d}{\rm{d} y} \big[N \fsnare(y)  + \vfusion(y) \big]+ \sqrt{2\eta\,\kb T}\,\xi(t),
\end{equation}
where, the energy \( \vsnare(N_{c},y) \) appearing in \eqref{eq:dynamics_y}---which depends on \( N_{c} \) and \( y \)---is replaced by the equilibrium free energy $\fsnare(y)=-\kb T \log\,\{
	\exp\,[-\vc(y)/(\kb T)\,] + \exp\,[-\vn(y)/(\kb T)\,] \,\} $---which depends only on  \( y \). This free energy is illustrated in Fig.~\ref{fig:results}[(D) dashed line].
The overall potential $\Phi(y)= N \fsnare(y)+\vfusion(y)$, driving the effective dynamics (\ref{eq:effective_langevin}) is shown by the solid red line in Fig.~\ref{fig:results}(D). It exhibits two local minima representing two metastable states.
The first metastable state [point $A$ in Fig.~\ref{fig:results}(D)] is located at \( y\simeq a \), where on average all the SNAREpins are in state $n$.
The second one  [point $B$ in Fig.~\ref{fig:results}(D)] is located at \( y_{f}<y_{2}<y_{*} \) and represents the  \emph{intermediate} state where on average all the SNAREpins are in state $c$, still confronting a reduced fusion barrier. 

The system evolving   in this energy landscape from the initial---\( y\simeq a \)---to the final---\( y=\yf \)--- state faces two successive  energy barriers $\Delta\Phi_1$ and $\Delta\Phi_2$.  With each barrier $\Delta\Phi_{1,2}$ one can associate a waiting time $\bar\tau_{1,2}$, that can be approximated by the Kramers formula \cite{Kramers_1940, Risken_1988,Schuss_2010} 
\begin{equation}
	\label{eq:kramers}
	\bar\tau_{1,2} = \tau_{\eta}\alpha_{1,2} \exp\left[\Delta\Phi_{1,2}/(\kb T)\right].
\end{equation}
The values of the numerical prefactors \( \alpha_{1,2} \) are determined by the local curvatures of the potential \( \Phi \) at its critical points and depend weakly on $N$, see Materials and Methods~\ref{sub:adiabatic_elimination_of_the_variable_boldsymbol_n__c}.

The approximated timescales $\bar\tau_{1,2}$  are compared with the  numerically computed values $\tau_{1,2}$  in Fig.~\ref{fig:results}[(C) solid lines]. 
The excellent agreement between the two sets of results suggests that the whole fusion process can effectively be described by two successive `chemo-mechanical' reactions, and that the rates in \eqref{eq:chemical_reaction} can be computed from the formulas \( k_{1,2} = \bar\tau_{1,2}^{-1} \)   while the remaining rate \( k_{-1} \) is prescribed by the condition of detailed balance.

Finally, note that with the parameters reported in Table~\ref{tab:parameters}, \( k\tau_{\eta}=4.5 \), which shows that our effective model is accurate even if the condition \( k\tau_{\eta}\gg 1 \) is not fully satisfied.

The peculiar dependencies of the waiting times $\tau_{1,2}$ on the number of SNAREpins $N$ can be now  understood  by  referring to  the  $N$-dependence of the energy barriers \( \Delta\Phi_{1,2} \).

\paragraph{ Timescale \( \boldsymbol{\tau_{2}} \): The cooperative action of the SNAREs reduces the time for crossing the fusion barrier.} 

In the intermediate state [point $B$ in  Fig.~\ref{fig:results}(D)], the SNAREpins are all in  state $c$ and the pulling force they apply on the membranes is exactly balanced by the short range repulsive forces.
The system remains trapped in this state until a thermal fluctuation delivers the energy \( \Delta\Phi_{2} \) allowing the system to reach the distance \( y=y_{f} \), where the fusion occurs.

In the absence of SNAREs, this energy difference is simply the bare fusion barrier $\vf$, see Fig.~\ref{fig:model}(D).
When the SNAREpins are present, the total force they apply brings the two membranes in close contact, which reduces the energy barrier.
This effect is amplified by an increase in  the number of SNAREpins: the larger  the number of SNAREpins, the larger the overall force, so the closer the membrane can be brought together, see Fig.~\ref{fig:results}(D).

Since the inter-membrane potential  \(\vfusion(y)\) decays rapidly as \( y \) increases, we can approximate the second energy barrier by \(\Delta\Phi_{2}\simeq \vf-Nw \), where \( w \) represents the amount of mechanical work that a single SNAREpin can deliver, see Materials and Methods~\ref{sub:estimation_of_the_mechanical_work_w} for the derivation of this result and the mathematical expression of \(w\).
According to \eqref{eq:kramers}, we then have 
\begin{equation}
	\label{eq:tau_two}
	\bar\tau_{2} (N) \propto \exp\left[-Nw/(\kb T)\right],
\end{equation}
hence the exponential decay of the time \( \tau_{2} \) with the number of SNAREpins.

With the parameters of Table~\ref{tab:parameters}, each SNAREpin provides a mechanical work \( w\simeq \SI{4.5}{\kbt}\) when it encounters the fusion barrier, which reduces the average time for fusion \( \bar\tau_2\) by a factor of \(\sim 100\), see Fig.~\ref{fig:results}(D).
This multiplicative effect allows fast fusion at the sub-millisecond timescale with as few as three SNAREpins.
For large enough number of SNAREpins (here \(N>7\)), the overall applied force  surpasses the membrane repulsion and the remaining fusion barrier disappears.
The obtained exponential decay of the timescale \( \tau_{2} \) with the number of SNAREpins suggest that the fusion could in principle proceed much faster than \( \SI{\sim 100}{\micro\second} \), being only limited by viscous forces.
Considering that each vesicle can accommodate up to \( \num{\sim100} \) SNAREpins, one cannot rule out the possibility of neurotransmitter release occurring much faster than \SI{100}{\micro\second}.
 Next we argue that such scenario is unlikely by showing that the fusion process gets slowed down if the number of SNAREpins becomes too large.

\paragraph{Timescale \( \boldsymbol{\tau_{1}} \): Increasing the number of SNAREpins slows down the synchronous zippering.} 

The average time \(\bar\tau_1\), taken for all the SNAREpins to switch from the \( n\) to the \(c\) conformation and then pull the membranes toward the bottom of the fusion barrier, exponentially increases with the number of SNAREpins \(N\), see Fig.~\ref{fig:results}C.
This dependence can be explained as follows.

As long as \( y_{*}<y<a\), the individual transition rates are such that \( k_+<k_- \) which implies that, on average, all the SNAREpins are in state \( n \) and therefore under compression, see Fig.~\ref{fig:model}(C). This idea is in agreement with the experimental results from Ref.~\cite{Zhang:2017fm}, that revealed the presence of the half-zipped metastable state.
Consequently in the interval \( y_{*}<y<a\) the average force \(- \rm{d}\fsnare/\rm{d}y\), that the SNAREpins collectively exert on the membranes, is repulsive.
Beyond the point \( y\simeq y_{*} \), the state \( c \) is stabilized (\( k_{+}>k_- \)) and the average force becomes attractive.
Since this force is proportional to the number of SNAREpins, the waiting time before a fluctuation can provide enough energy to surpass the repulsion---and overcome the barrier \( \Delta\Phi_{1} \)---increases with \( N \).
This constraint results from the mechanical feedback
induced by the membranes. The latter  play the role of a rigid backbone that forces  the SNAREpins to bridge approximately the same intermembrane distance, see Refs.~\cite{Caruel:2015im,Caruel:2017eba,Caruel:2018jg}.

To specify the \( N \)-dependence of \( \tau_{1} \), we use the fact that for \( y>y_{*} \) we can consider that \( \vfusion=0 \), so \( \Delta\Phi_{1} \) can be approximated by \(\Delta\Phi_{1}\simeq N[\fsnare(y_{*})-\fsnare(a)] \simeq  \deltav-\kb T\log(2) \), where \( \deltav=e_{n}(y_{*})-e_0\). According to \eqref{eq:kramers}, we can then write
\begin{equation}\label{eq:tau_one}
\bar\tau_{1}(N)\propto \exp\left[N\Delta e/(\kb T)\right],
\end{equation}
which shows that the timescale \( \bar{\tau_{1}} \) increases exponentially with the number of SNAREpins.

 \begin{figure}[tbp]
 	\centering
 	\includegraphics[scale=0.4]{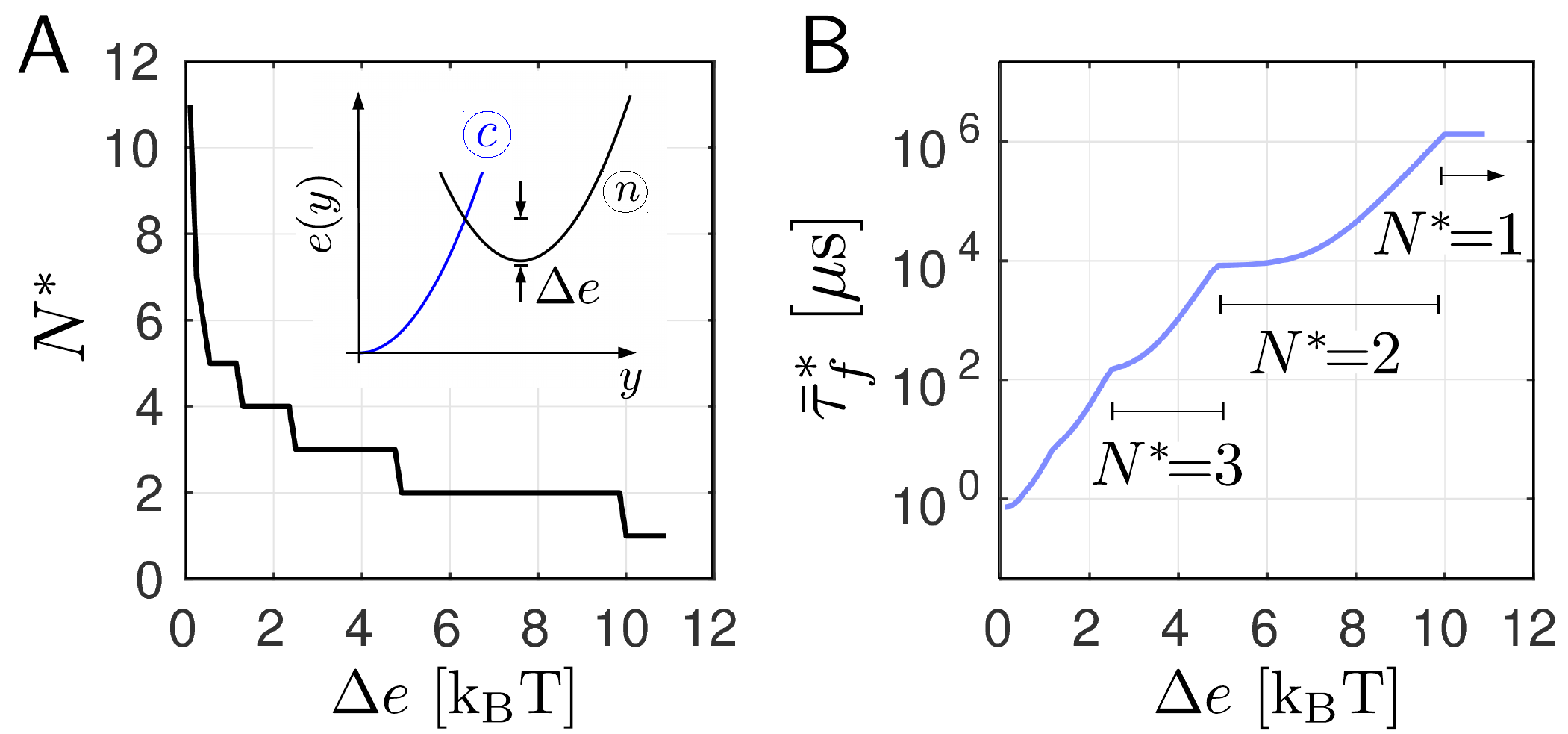}
 	\caption{
	Effect of the intrinsic energy barrier \( \deltav \) on the optimal number of SNAREpins (A) and on the associated fusion time (B).The parameters values are taken from Table~\ref{tab:parameters} with \( \kn = \SIrange[range-phrase = -,range-units = single]{0.11}{24}{\pico\newton\per\nano\meter}\).
 	}
 	\label{fig:EffectOfDeltaV}
 \end{figure}

From \eqref{eq:tau_one} we obtained that the first energy barrier is fully controlled by a single parameter  \( \deltav \) which therefore has a strong influence on both the existence and value of the optimal number of SNAREpins.
To study the  effect  of $\deltav$ on the fusion time, we varied the parameter \( \kn \)  describing  the curvature of the energy \( e_{n} \). The results of our parametric study are summarized in Fig.~\ref{fig:EffectOfDeltaV}. 
These data were obtained by using \eqref{eq:kramers} to compute the intersection of the curves \( \bar\tau_{1,2}(N) \) for each value of \( \deltav \).
We checked that the  results are in good agreement with  direct numerical simulations. Despite the broadness of the interval of parameter values tested, the optimal number of SNAREpins remains below \num{10}.
If we consider only the cases corresponding to sub-millisecond fusion times, we obtain \( N\geq 3 \) with \( \deltav<\SI{4}{\kbt} \). 
The latter value is compatible with recent estimate of \( \deltav\approx \SI{5}{\kbt} \) for the \( n\to c \) transition energy barrier, see Refs.~\cite{Li:2016isa,Zhang:2017fm}.
Note also that the predicted optimal  number of SNAREpins   is robust  because it corresponds to a  \emph{plateau}  on the  \( N_{*}(\deltav )\) curve, see Fig.~\ref{fig:EffectOfDeltaV}(A).



\paragraph{Robustness of the predictions. } 

\begin{figure}
	\centering
	\includegraphics[scale=0.4]{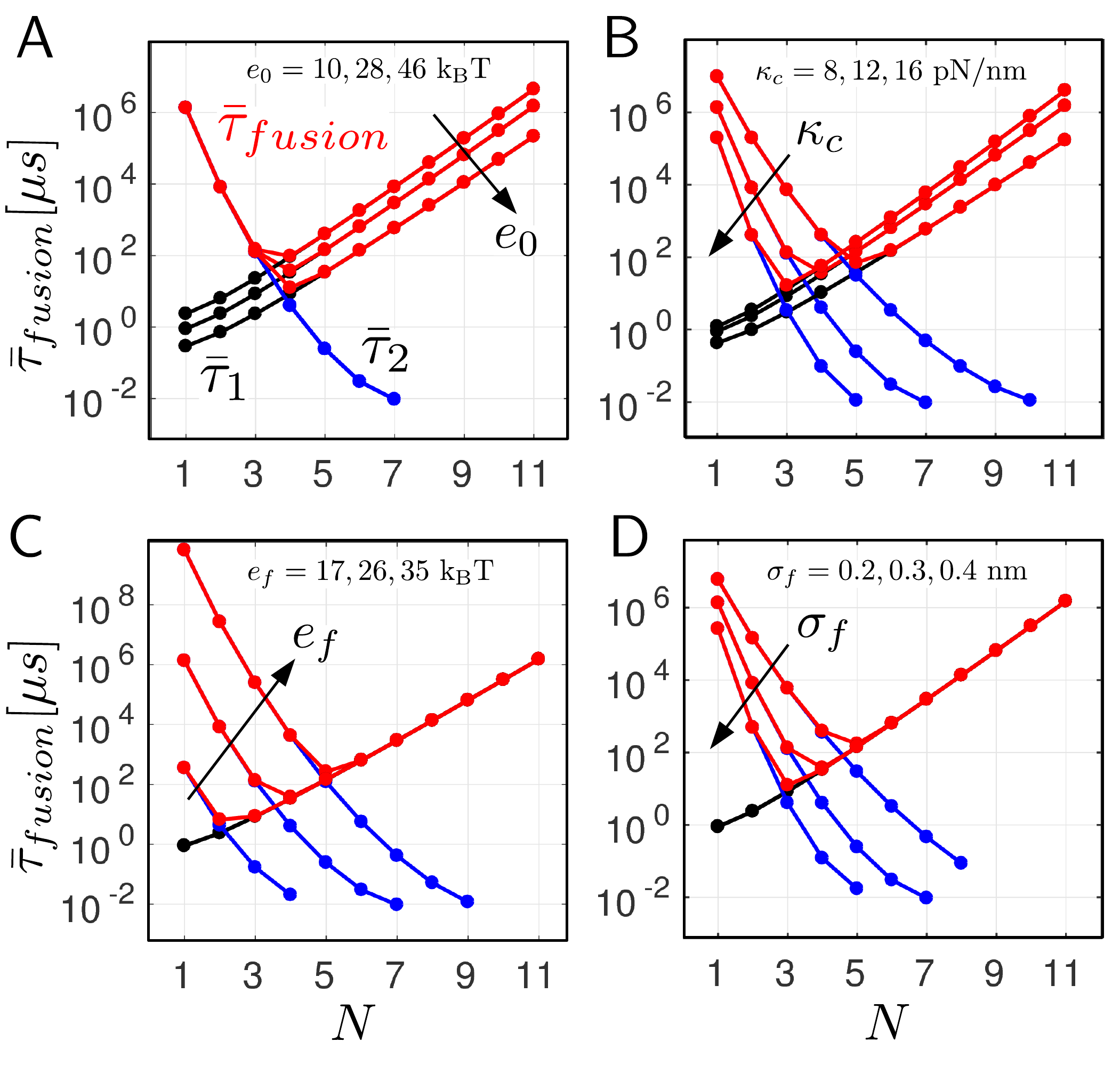}
	\caption{
	Robustness of the prediction. Influence of the parameters \( \vzero \) (A), \( \kc \) (B), \( \vf \) (C) and \( \sigmaf \) (D), on the timescales \( \bar{\tau}_{1,2} \) and \( \taufusionbar \).
	The results were obtained using \eqref{eq:kramers}.
	}
	\label{fig:Robustness}
\end{figure}

 The results presented above, were obtained for the  parameters values listed in Table~\ref{tab:parameters}. For some of these parameters only a rough estimate  is available  at this stage, see SI Appendix. 
To test the robustness of our theoretical predictions, we computed the average waiting times $\bar\tau_{1,2}(N)$ from \eqref{eq:kramers} for different values  of four key parameters of the model, $\vzero$, $\kc$, $\vf$, $\sigmaf$, see Fig.~\ref{fig:Robustness}.
For each of these parameters, the lower and the upper bounds delimit broad intervals covering the values obtained from different experimental studies.

A comparison between Fig.~\ref{fig:Robustness} and Fig.~\ref{fig:results}(B) shows that our  results are  only marginally affected by changes in the parameter values. In particular, the existence of a sharp minimum of the fusion time, associated with an optimal number of SNAREpins, is a robust prediction.
In addition, the value of the optimal number of SNAREpins, is weakly sensitive to the parameters: it always remains in between $3$ and $6$.
Remarkably, despite the large difference between the upper and lower bounds for each of the parameters, the average fusion time remains in the sub-millisecond scale.

The energy landscape associated with the zippering of the SNARE complexes is the object of intense current research   \cite{Li:2016isa,Walter:2010dt,Zhang:2017fm}. In our model this landscape is fully characterized by only four parameters: the distance \( a \), the energy  \( \vzero \), and the stiffnesses \( \kappa_{n,c} \).
While the distance \( a \) has been measured with precision in recent works \cite{Gao:2012gu,Zhang:2017fm}, the values of the other three parameters are still not known with certainty.  Several estimates of the energy bias \( \vzero \) lying between \num{20} and \SI{40}{\kbt} can been found in the literature \cite{Zhang:2017fm,Li:2007ka}. We show in Fig.~\ref{fig:Robustness}(A) that variations within this interval  affect  mostly \( \tau_{1} \), change the fusion time by one order of magnitude,  but have  almost no effect on the optimal number of SNAREpins. 
Currently, only indirect evaluation of the stiffnesses \( \kappa_{n,c} \) can be obtained from the available data, see SI Appendix.  Within the broad range  of values tested in our numerical simulations we again observed only small variations of the optimal number of SNAREpins, see Fig.~\ref{fig:EffectOfDeltaV}, and Fig.~\ref{fig:Robustness}(B).

One of the most documented physical phenomena involved in the fusion process is the merging of the two membranes. The amplitude of the associated repulsion force  depends, in our model, on the parameters \( \vf \) and \( \sigmaf \), whose influence  on the fusion time is illustrated in Fig.~\ref{fig:Robustness}[(C) and (D)], respectively.
As expected from the analysis presented in Section~\ref{sec:results}, changing the values of these two parameters   affect only the height of energy barrier \( \Delta\Phi_{2} \) and therefore the timescale \( \tau_{2} \).
Increasing $\vf$ raises the height of the maximum of $\vfusion$ [see Fig.~\ref{fig:model}(D]), while decreasing $\sigmaf$ deepens the second energy well [point B in Fig.~\ref{fig:results}(D)], which results in both cases in the increase of $\tau_2$.
This leads \emph{in fine} to the increase of the optimal number of SNAREs.
Notice that $\vf$ depends on the type of lipids, on the membrane curvature and is also strongly sensitive to the membrane tension \cite{markvoort2011,finkelstein1986,grafmuller2007,grafmuller2009,lee2008}. Therefore, its value can be different in different cells or experimental set-ups. In particular, we expect the \emph{in vivo} value to be smaller than the value measured in artificial systems (\SI{35}{\kbt} ), which in general use low tension and low curvature membranes, see Ref.~\cite{FrancoisMartin:2017db}.

In conclusion, while additional experimental studies are needed to refine the calibration of the  model, the above parametric study shows the robustness of the effects of the mechanical crosstalk between the SNAREpins. 


\section{Discussion} 
\label{sec:discussion}

In this paper, we elucidated the central role played by mechanical coupling in synchronizing the activity of  SNAREpins, which is necessary to enable sub-millisecond release of neurotransmitters.
Our approach to the problem   complements previous studies focussed predominantly on the molecular details of the single SNARE zippering transition \cite{markvoort2011,risselada2012,Walter:2010dt,Acuna:2014km,Hernandez:2012cg,Li:2014dg,Sutton:1998cw,Stein:2009jy, Gao:2012gu,Zhang:2017fm}.

As a starting point we used  a previously unnoticed analogy between the activity of SNARE complexes  and the functioning of myosin II molecular motors.
Viewed broadly, both systems ensure ultra-fast mechanical contraction.
In the case of muscle, destabilization of the  pre-power stroke state is the result of a mechanical bias created by an abrupt shortening of the myofibril \cite{Huxley_1971, Ford_1977}. In the case of SNAREs, similarly abrupt destabilization is a result of the calcium-induced removal of the synaptotagmin-based clamp, most likely when \ce{Ca^2+} triggers disassembly of the synaptotagmin ring \cite{Ramakrishnan:2019}.

To pursue this analogy, we developed a variant of the power-stroke model of  Huxley and Simmons ~\cite{Huxley_1971,Caruel:2016kw}, in which the zipping is viewed as a transition between two discrete states endowed with different elastic properties.
This representation is supported by recent experiments \cite{Gao:2012gu,Zhang:2017fm}, which provided essential data for the calibration of the model. 

Our analysis of the collective behavior of $N$ ``switchers'' of this type suggests that the main function of the SNARE machinery is to bring the two membranes to a distance beyond which the fusion process can proceed spontaneously.
The emerging intermediate configuration, where the two membranes are sufficiently closely tethered, can be then viewed as an intermediate state in the reaction process linking the fused and unfused states.
The result is a representation of the SNARE mediated fusion as a two stage reaction.

We linked the first stage of the process with the collective zippering of the SNAREpins and showed that this step gets exponentially more sluggish as the number of SNAREpins increases. This phenomenon was studied previously in the context of muscles, see Refs.~\cite{Caruel:2013jw,Caruel:2018jg}. It originates  (i) from the experimentally suggested presence of a metastable half-zipped state along the zippering free energy landscape \cite{Zhang:2017fm}, and (ii) from the long-range mechanical interactions mediated by the scaffolding membranes, which create a negative feedback preventing a fast collective escape from the metastable half-zippered state.

The second stage of the process is the transition from the intermediate state to the fused state. The associated timescale \( \tau_{2} \) decreases exponentially  with the number of SNAREpins because the larger the number of acting SNAREpins, the closer the membranes can be brought together in the intermediate state and therefore the higher is the energy of this state. This results in an exponential decay of the timescale \( \tau_{2} \) with the number of SNAREpins. 
Behind this phenomenon is the presence of a residual force in the configuration where the SNAREpins have reached the intermediate state. This perspective is supported by the results of Ref.~\cite{Gao:2012gu,Zhang:2008bp}.

The antagonistic \( N \)-dependence of the rates characterizing the two stages reveals the existence of an optimal number of SNAREs \( N_{*} \) allowing the system to perform fusion at the physiologically appropriate timescales. Our prediction \( N_{*}=\numrange{4}{6} \) is supported by recent in situ cryoelectron microscope tomography observation, see Ref.~\cite{Li:2019}.

We remark that our result strongly depend on the initial configuration of the system, which we link with the structure of the fusion machinery immediately after Synaptotagmin removal by calcium.
Notice that the position \( y_{*} \) of the barrier separating the half-zippered and the fully zippered states is such that \( y_{*}<a \). Therefore the timescale \( \tau_{1} \) exists only if the initial membrane separation \( y_{0}>y_{*} \).
This assumption seems to be supported by experiments, see Ref.~\cite{Gruget:2018dr}.
It has previously been reported that, upon approach of two membranes devoid of SNAREs, Synaptotagmin exerts repulsive force from \SI{10}{\nano\meter} down to \SI{4}{\nano\meter} where it becomes a repulsive wall \cite{Gruget:2018dr}. According to this result, \( y_{0} \) should range between \num{4} and \SI{10}{\nano\meter}. However it is probably slightly larger  under physiological conditions because of the presence of the SNAREs. 
With the parameters adopted in our simulations (see Table~\ref{tab:parameters}), the position of the barrier is \( y_{f}\simeq \SI{4.5}{\nano\meter} \) in accordance with Refs.~\cite{Gao:2012gu,Zhang:2017fm}, see Fig.~\ref{fig:results}. Therefore, in all likelihood, \( y_{0} \) is larger than \( y_{*} \) and our predictions should be valid.

Finally we mention the fact that the timescale \( \tau_{2} \) exponentially decrease with \( N \) seems to be supported by experimental studies reporting sub-millisecond fusion time with \( N=\numrange{3}{6} \) \cite{Hua:2001bb,Sinha:2011gxa,Mohrmann:2010eg}
However, in a recent theoretical study the decay was also found to be exponential but with a much slower decay: the cooperation of at least sixteen SNAREs was predicted to be necessary to reach the physiological fusion time \( \SI{\sim 100}{\micro\second} \) \cite{Mostafavi:2017gd,McDargh:2018jd}. The difference is explained by the fact that the residual work in this study is \(w=\SI{0.48}{\kbt}\) instead of \SI{4.5}{\kbt} in our model, see\eqref{eq:tau_two}.
This difference originates from the assumption made by the authors that the zippering energy of a SNARE complex is entirely dissipated before the membranes encounter the fusion barrier. In other words, the authors have implicitly assumed that after the calcium entry, the zippering of the SNAREpins does not generate any pulling force to assist fusion and concluded that the remaining  residual force is of entropic nature.
Recent direct microscopic observations implying that synaptic fusion involves only six SNAREpins \cite{Li:2019} would appear to invalidate this assumption. 

In conclusion,  our model describes membrane fusion by a team of mechanically interacting SNAREpins as  a two stage process. We show that conventional  biochemical and  biophysical measurements cannot be used directly  to predict the 
 associated rates and that mechanical modeling is crucial for linking these rates with independently measured parameters. 
  Our work emphasizes the importance of identifying  mechanical pathways and  specifying mechanistic feedbacks.  The main conceptual  outcome of our study is the realization   that in the case of synaptic fusion,  SNARE proteins can  perform optimally only if they act collectively. 
  The remarkable fact is that when the team is of the optimal size,  such synchronization is not deterred by thermal fluctuations, which guarantees that the collective strike is  simultaneously fast, strong and robust. 

Finally, we mention that the synaptic fusion  is  only one  of many biophysical processes involving mechanically-induced collective conformational changes. Other examples include  ion gating   in   hair cells \cite{Martin_2000,Bormuth:2014hh}, collective decohesion of adhesive clusters \cite{Yao:2006de,Erdmann_2007}, folding-unfolding of macromolecular hairpins   \cite{Woodside:2008uz,Bosaeus:2012kp,Liphardt:2001fp} and  folding of ParB-ParS complexes in DNA condensation \cite{Chen:2015cv,Funnell:2016bo}.   
In each of these situations one can identify a dominating long-range mechanical interaction making  the  theoretical  framework  developed  in this paper potentially useful.

\matmethods{
\subsection{Rigid membrane assumption} 
\label{sub:rigid_membrane_assumption}
We assume for simplicity that the vesicle and the target membranes are rigid, which implies that all the SNAREpins share the same inter-membrane distance \( y \).
This approximation is valid if the characteristic length \( \ell \) associated with the deformation generated by a single SNAREpin is large compare to the size of the SNARE bundle.
We can use the following estimate \( \ell=\sqrt{\kappa/\sigma} \), where \( \kappa \) is the membrane rigidity and \( \sigma \) is the membrane tension.
We have typically \( \kappa\sim\SIrange{20}{50}{\kbt} \) and \( \sigma\sim\SIrange{e-04}{e-06}{\newton\per\meter} \) so \( \ell\sim\SIrange{30}{120}{\nano\meter} \). Since the size of the SNARE bundle is less than \SI{10}{\nano\meter}, our assumption should be valid.

\subsection{Model of a single SNAREpin} 
\label{sub:model_of_a_single_snarepin}
We set, for simplicity, that the energies \( \vn \) and \( \vc\) of the SNAREpins in the states $n$ and $c$, respectively, depend on $y$ quadratically, so that
\begin{equation}
	\label{eq:en_ec}
\begin{split}
	\vn(y) &= (\kn/2)(y-a)^{2} + \vzero,\\
	\vc(y) &= (\kc/2)y^{2},
\end{split}
\end{equation}
where \( \kappa_{c,n} \) represent lumped stiffnesses parameters. 
We denote \( y* \) the distance where \( \vn(y_{*})=\vc(y_{*}) \), see Fig.~\ref{fig:model}(C).
In the absence of external load (zero force), the stable states are located at \( y=a \) and \( y=0 \). In this situation the entire energy associated with the zippering process is consumed when the SNAREpin reaches state \( c \) at \( y=0 \), which can then be considered as a ground state with zero energy.

The rates \( k_{\pm} \) of the \( n \rightleftharpoons c \) transitions obey the detailed balance relation $k_{+}/k_{-} = \exp\left[(\vc -\vn)/(\kb T)\right]$, with the bias towards the direct  transition  $n\to c$, i.e. \( k_+ > k_-\),     at \( y < y_{*} \) and conversely,  in the direction of the reverse transition $c\to n$ at \( y > y_{*} \), see Fig.~\ref{fig:model}(C)

For simplicity, and following \cite{Huxley_1971}, we consider that the transition from the high to the low energy state occurs at a constant rate $k$, which fixes the characteristic timescale of the conformational change. This assumption could be  easily 
replaced by a more adequate one
 at the expense of introducing two additional parameters but with only a minimal impact on the results, see Ref.~\cite{Caruel:2017eba}.
With this assumption and using the detailed balance, we write the transition rates as
\begin{equation*}
\begin{split}
k_-(y)=k,&\; k_+(y)=k\exp\{[\vn(y)-\vc(y)]/(\kb T)\},\;{\rm if}~ y>y_{*}\\ 
k_+(y)=k,&\; k_-(y)=k\exp\{[\vc(y)-\vn(y)]/(\kb T)\},\;{\rm if}~ y<y_{*}.\\ 
\end{split}
\end{equation*}


\subsection{Numerical implementation of the model} 
\label{sub:numerical_implementation_of_the_model}
The discrete stochastic process associated with the variable \( N_{c} \) was simulated as a two-state Markov chain with a fixed timestep \( \Delta t=10^{-6}\,t_{\eta} \). At each timestep the transition probabilities \( W_{+1,-1,0}\Delta t \) are computed and the next event is chosen based on an acceptation-rejection condition using a random number uniformly distributed between 0 and 1.
The Langevin equation was simulated using a first order explicit Euler scheme.
More details about the computer algorithms can be found in SI Appendix.

\subsection{Adiabatic elimination of the variable \( \boldsymbol{N_{c}} \)} 
\label{sub:adiabatic_elimination_of_the_variable_boldsymbol_n__c}
We consider the situation where \( t_{\eta}\gg k^{-1} \): the characteristic time of the conformational changes is negligible compared to the timescale associated with the relaxation of the vesicle's position. In this limit, the conformational state of each SNAREpin can be considered at equilibrium. Therefore for a given position of the vesicle \( y \), the probability of a configuration with \( N_{c} \) SNAREpins in state \( c \) follows the Boltzmann distribution
\begin{equation}
	\label{eq:distrib_nc}
\rho(N_{c};y)=\frac{1}{Z(y)}\binom{N}{N_{c}}\exp
\left\{
-\left[
N_{c}\vc(y)+(N-N_{c})\vn(y)
\right]/\left(\kb T\right)
\right\},
\end{equation}
where \( \binom{N}{N_{c}}=\frac{N!}{N_{c}!(N-N_{c})!} \).
We then integrate \eqref{eq:dynamics_y} with respect to the distribution (\ref{eq:distrib_nc}) and obtain \eqref{eq:effective_langevin}. Since the energy \( \vsnare \) is linear in \( N_{c} \), our approximation results in replacing \( N_{c} \) by its average \( \langle n_{c}\rangle(y)=\sum_{N_c}N_{c}\rho(N_{c};y) \) in \eqref{eq:dynamics_y}.
In \eqref{eq:kramers}, the prefactors are given by \cite{Schuss_2010} \( \alpha_{1,2} = \frac{2 \pi \kb T} {a^2 \sqrt{\Phi_{1,2}''(y_{\mbox{\tiny max}}) |\Phi_{1,2}''(y_{\mbox{\tiny min}})|}} \), where \( y_{\mbox{\tiny max}} \) and \( y_{\mbox{\tiny min}} \) denote the positions of the considered barrier and minimum, respectively.

\subsection{Estimation of the mechanical work \( \boldsymbol{w} \)} 
\label{sub:estimation_of_the_mechanical_work_w}
In the intermediate state, inter membrane distance \( y_{2} \) is sufficiently lower than the threshold \( y_{*} \), so that the free energy can be well approximated by the energy of the state \( c \). We then write \( \Phi(y)\simeq \vfusion(y)+N\frac{\kc}{2}y^{2} \) which leads to the following expression for the energy barrier separating the intermediate state and the fused state,
\[ 
\Delta\Phi_{2}=\vf\left\{(1-\exp\left[-(y_{2}-y_{f})^{2}/(2\sigmaf^{2})\right]\right\} + N\frac{\kc}{2}(\yf^{2}-y_{2}^{2})
.\]
By noting that \( y_{2} \) verifies \( \left.\frac{\rmd\Phi(y)}{\rmd y}\right|_{y=y_{2}}=0 \), we obtain
\(
\Delta\Phi_{2}=\vf - N w,
\)
with \[ w\simeq\kappa_{c}\big(y_{2}^{2}-\yf^{2}+\frac{\sigmaf^{2} y_{2}}{y_{2}-\yf}\big)\geq0. \]
 Notice that since the energy \( \vfusion \) decays rapidly for \( y>\yf \), the parameter \( y_{2} \) depends weakly on \( N \).
}
\showmatmethods{}

\acknow{We would like to thank Yongli Zhang  for providing us with the data used for the calibration of our model and Ben O’Shaughnessy for stimulating discussions.
This work was supported by a European Research Council (ERC) funded grant under the European UnionQs Horizon 2020 research and innovation programme (grant agreement no. 669612) to J.E.R
}

\showacknow{}

\bibliography{references}

\end{document}